\documentclass[useAMS,usenatbib]{mn2e}
\usepackage{graphicx,graphics,amsmath,amssymb,upgreek}
\usepackage[T1]{fontenc}
\usepackage{aecompl}
\usepackage{url}

\title[Fast Radio Bursts at 145 MHz with ARTEMIS]
{Limits on Fast Radio Bursts at 145 MHz with ARTEMIS, a real-time
  software backend} 
\author[Karastergiou et al.]{A.~Karastergiou$^{1,2,3}$,
J.~Chennamangalam$^1$, 
W.~Armour$^4$,
C.~Williams$^1$,
B.~Mort$^4$,
\newauthor  
F.~Dulwich$^4$,
S.~Salvini$^4$, 
A.~Magro$^5$,
S.~Roberts$^6$,
M.~Serylak$^3$,
A.~Doo$^7$,
A.~V.~Bilous$^{8}$,
\newauthor   
R.~P.~Breton$^{9}$,
H.~Falcke$^{8,10}$,
 J.-M.~Grie\ss{}meier$^{11, 12}$,
J.~W.~T.~Hessels$^{10, 13}$, 
E.~F.~Keane$^{14}$,  
\newauthor  
V.~I.~Kondratiev$^{10, 15}$
M.~Kramer$^{16}$,  
J.~van Leeuwen$^{10, 13}$,
A.~Noutsos$^{16}$, 
S.~Os{\l}owski$^{16, 17}$,
\newauthor  
C. Sobey$^{10}$,
B.~W.~Stappers$^{9}$, 
P.~Weltevrede$^{9}$,
\\
$^1$Astrophysics, University of Oxford, Denys Wilkinson Building,
Keble Road, Oxford OX1 3RH, UK\\
$^2$Department of Physics and Electronics, Rhodes University, PO Box
94, Grahamstown 6140, South Africa\\
$^3$Physics Department, University of the Western Cape, Cape Town
7535, South Africa\\
$^4$Oxford e-Research Centre, University of Oxford, Keble Road, OX1
3QG, United Kingdom\\
$^5$Institute of Space Sciences and Astronomy, University of Malta,
Msida. MSD2080. Malta\\
$^6$Information Engineering,University of Oxford, Oxford OX1 3PJ, UK\\
$^7$STFC Chilbolton Observatory\\
$^{8}$Department of Astrophysics/IMAPP, Radboud University Nijmegen, PO Box 9010, 6500 GL Nijmegen, The Netherlands\\
$^{9}$Jodrell Bank Centre for Astrophysics, School of Physics and Astronomy, The University of Manchester, Manchester M13 9PL, UK\\
$^{10}$ASTRON, The Netherlands Institute for Radio Astronomy, Postbus 2, 7990 AA Dwingeloo, The Netherlands\\
$^{11}$LPC2E - Universit\'{e} d'Orl\'{e}ans/CNRS\\
$^{12}$Station de Radioastronomie de Nan\c{c}ay, Observatoire de Paris - CNRS/INSU, USR 704 - Univ. Orleans, OSUC, 18330 Nan\c{c}ay, France\\
$^{13}$Anton Pannekoek Institute for Astronomy, University of Amsterdam, Postbus 94249, 1090 GE Amsterdam, The Netherlands\\
$^{14}$SKA Organisation,  Jodrell Bank Observatory, Lower Withington,
Macclesfield, Cheshire, SK11 9DL, UK\\
$^{15}$Astro Space Centre, Lebedev Physical Institute, Russian Academy of Sciences, Profsoyuznaya Str. 84/32, Moscow 117997, Russia\\
$^{16}$Max-Planck-Institut f\"{u}r Radioastronomie, Auf dem H\"{u}gel 69, 53121 Bonn, Germany\\
$^{17}$Fakult\"{a}t f\"{u}r Physik, Universit\"{a}t Bielefeld, Postfach 100131, D-33501, Bielefeld, Germany\\
 } \date{\today}
\begin{document}
\pagerange{\pageref{firstpage}--\pageref{lastpage}} 
\maketitle
\label{firstpage}
\begin{abstract}
  Fast Radio Bursts (FRBs), are millisecond radio signals that exhibit
  dispersion larger than what the Galactic electron density can
  account for.  We have conducted a 1446 hour survey for Fast Radio
  Bursts (FRBs) at 145~MHz, covering a total of 4193 sq. deg on the
  sky. We used the UK station of the LOFAR radio telescope -- the
  Rawlings Array -- , accompanied for a majority of the time by the
  LOFAR station at Nan\c{c}ay, observing the same fields at the same
  frequency. Our real-time search backend, ARTEMIS, utilizes graphics
  processing units to search for pulses with dispersion measures up to
  320 cm$^{-3}$ pc.  Previous derived FRB rates from surveys around
  1.4~GHz, and favoured FRB interpretations, motivated this survey,
  despite all previous detections occurring at higher dispersion
  measures. We detected no new FRBs above a signal-to-noise threshold
  of 10, leading to the most stringent upper limit yet on the FRB
  event rate at these frequencies: 29 sky$^{-1}$ day$^{-1}$ for
  5~ms-duration pulses above 62~Jy. The non-detection could be due to
  scatter-broadening, limitations on the volume and time searched, or
  the shape of FRB flux density spectra.  Assuming the latter and that
  FRBs are standard candles, the non-detection is compatible with the
  published FRB sky rate, if their spectra follow a power law with
  frequency ($\propto \nu^{\alpha}$), with $\alpha\gtrsim+0.1$,
  demonstrating a marked difference from pulsar spectra. Our results
  suggest that surveys at higher frequencies, including the low
  frequency component of the Square Kilometre Array, will have better
  chances to detect, estimate rates and understand the origin and
  properties of FRBs.
\end{abstract}


\section{Introduction}

High-time resolution radio astronomy is entering a new era due to two
developments. Firstly, new radio telescopes are being built that
combine very high instantaneous sensitivity with wide fields of view,
as required for blind searches of short duration events of transient
radio emission. Secondly, high performance computing (HPC) enables us
to process the data extremely quickly, providing the opportunity to
discover and react to such events, enabling localization and
classification.

Our quest to explore the dynamic nature of the radio sky is
underpinned by the realization that bright, short duration radio
bursts must originate from extreme physical processes taking place
around compact sources. Neutron stars offer prime examples of such
phenomena, being reliable emitters of short duration (typically 50~ms
or less) radio pulses \citep{lor05}. These signals are observed once
per rotational period for the majority of ordinary pulsars, but can be
as infrequent as a few pulses per day in the case of Rotating Radio
Transients \citep[RRATS;][]{mcl06}.  In addition to neutron stars, a
new class of extraordinary sources has been identified in high time
resolution radio astronomy which highlights the importance of
real-time processing. The first of these sources -- that have now come
to be known as Fast Radio Bursts (FRBs) -- was discovered by
\cite{lor07}, who reported the detection of a burst using the Parkes
radio-telescope, with evidence supporting large, extragalactic
distance.  Following this discovery, more such bursts have been
detected \citep{kea12,tho13,spi14,bur14, pet15,rav14}. We stress that
all these events except \cite{pet15} were discovered via offline
processing of recorded data, therefore no immediate follow-up
observations to localize them were possible.  The origin of FRBs is
unclear. Proposed explanations include flaring magnetars
\citep{pop13}, binary neutron star mergers \citep{tot13},
gravitational collapse of neutron stars to black holes \citep{fal14},
emissions from companions of extragalactic pulsars \citep{mot14}, and
at least one Galactic proposition in the form of nearby flare stars
\citep{loe14}. These associations can be strengthened or weakened by
comparing the rates of these events to the observed rates of
FRBs. Current best limits from the observed FRBs are given by
\cite{tho13} as 10$^4$~sky$^{-1}$~day$^{-1}$, or a volumetric rate of
10$^{-3}$~gal$^{-1}$~year$^{-1}$.

In general, an FRB will be dispersed by the free electron content of
the medium it propagates through. As a result, the lower frequency
components of a broadband burst arrive later than their high frequency
counterparts, the delay at a given frequency $\nu$ being proportional
to the line integral of the electron column density along the line of
sight (known as the dispersion measure) and $\nu^{-2}$. There is some
evidence that the dispersion measure (DM) of one of the aforementioned
FRBs -- the \cite{kea12} burst -- may be explained by the Galactic
distribution of electrons \citep{ban14}.  If FRBs are indeed
extragalactic in origin, they can be used to quantify the ionized
matter in the intergalactic medium (IGM), thereby allowing us to
determine the baryon content of the universe. This could help solve
the long-standing `missing baryon problem' in cosmology, where there
is a discrepancy between the observed and the expected quantities of
baryons \citep{per92,mcq14}. Answering this question is one of the
science objectives of the Square Kilometre Array \citep[SKA][]{mac15}.

There is an expectation that astrophysical bursts will emit across a
broad range of radio frequencies.  The observed signal-to-noise ratio
of an FRB can be maximized by correcting for the dispersion induced
delays across the frequency band of a telescope, in a well established
process from pulsar astronomy called dedispersion. In the specific
context of a blind search for FRBs, where the DM is not known, the
main challenge is to perform the incoherent dedispersion transform on
the data. By this, we mean the process of producing from time series
of total power in a large number of narrow frequency channels, time
series of total power over a large range of DM values, revealing
potential FRB candidates. This is a computationally expensive task,
and performing it in real time requires specialized software running
on high performance computers with accelerated hardware. In the last
few years, implementations of the accelerated dedispersion transform
that are capable of real-time processing of typical data streams have
emerged \citep{arm12,bar12,mag11} and are gradually being applied to
surveys.

Comprehensive arguments in favour of searching for individual bright
and dispersed pulses have been present in the literature and applied
in practice for many years \citep[see, for example,][]{mcl03}. Yet
despite the discoveries mentioned above, we are still in the early
stages of putting together a complete picture of the dynamic nature of
the radio sky. The most up-to-date limits hitherto on fast transient
sources at 150~MHz can be found in \cite{coe14}, who use the
results of the LOFAR Pilot Pulsar Survey (LPPS) to derive an upper
limit on the FRB rate of 150 sky$^{-1}$ day$^{-1}$. All aforementioned
FRBs have been discovered at a frequency around 1400~MHz. Upper limits
to the FRB rate at different frequencies provide constraints on the
luminosity distribution and the spectral characteristics. Assuming the
flux density spectra of FRBs follow a power law, then
\begin{equation}
  S(\nu) \propto \nu^\alpha,
\end{equation}
where $S(\nu)$ is the flux density of the burst as a function of
frequency $\nu$, $\alpha$ is the spectral index. Low frequency FRB
surveys are also being carried out at the Murchison Widefield Array
(MWA), a low frequency dipole array located in Western Australia
operating between 80 and 300~MHz. \cite{tro13} suggest several
expected FRBs per observing week, providing a clear framework for
estimating FRB rates depending on their intrinsic spectral index and
scattering properties. There is currently no published detection of an
FRB from the MWA in the literature.

The Low Frequency Array, LOFAR, is currently the most sensitive radio
telescope in the world at frequencies between 30 and 250~MHz. It is a
prototypical example of a `next-generation radio telescope' and a
pathfinder towards the Square Kilometre Array (SKA). The telescope is
described in detail in \cite{van13}. Discovery of pulsars and fast
transients is one of the key science goals of LOFAR, as outlined in
\cite{sta11}. We present here details and first results of a programme
to use one international LOFAR station to search for FRBs, in an
effort to characterize the dynamic nature of the radio sky at LOFAR
frequencies. In this paper, we describe the system developed to
conduct fast transient surveys and the results of the first survey.

This paper is organized as follows. In \S\ref{sec_sysdesc}, we
describe the project, including details of the hardware and software,
and the signal processing algorithms, and in \S\ref{sec_surv}, we
describe our survey for fast transients using ARTEMIS, before
concluding in \S\ref{sec_concl}.

\section{The ARTEMIS project}\label{sec_sysdesc}

We have put together the hardware and software that is required to
continuously monitor the large LOFAR fields of view and detect FRBs in
real time using general purpose computing on graphics processing units
(GPUs).  The scientific potential and the technical challenge of
real-time FRB discovery were the prime motivating factors behind the
Advanced Radio Transient Event Monitor and Identification System
(ARTEMIS) project, with an emphasis on applying HPC techniques to data
from next generation telescopes. The primary goals of the ARTEMIS
project therefore consist of:
\begin{itemize}
\item Enabling the science goals of the SKA and its pathfinders in the
  area of surveys for new pulsars and exotic fast transients.
\item Exploiting many-core technologies to minimize capital costs of
  hardware, whilst delivering energy-efficient solutions for
  streaming processing in radio astronomy.
\end{itemize}

\subsection{Technical description}

We use the high-band antennas (HBA) of a single international LOFAR
station, which consists of 96 dual-polarization antennas covering a
frequency range between 110 and 250~MHz.  The ARTEMIS installation
currently uses four 12-core servers, each equipped with a single Fermi
NVIDIA GPU card located on site at the Rawlings Array in the UK (a
similar setup exists at the international LOFAR station in Nan\c{c}ay,
France and an ARTEMIS server for offline pulsar data processing exists
in J\"ulich, Germany). The servers receive data through a broadband
(10 Gbps) switch, which is also responsible for sending the data back
to the LOFAR correlator in the Netherlands, for normal operations of
the International LOFAR Telescope (ILT).  Each LOFAR station generates
a total of 3.2 Gbps of beamformed data, which corresponds to a
sky bandwidth of approximately 48~MHz using 16-bit sampling. The
station beamformer produces a total of 244 beamlets each with the
frequency and time resolution specified in Table
\ref{tab:specs}. Estimated values of the system equivalent noise are
also given, for an observation with a total bandwidth of 6~MHz and
time resolution of 327.68~$\upmu$s, by scaling the sensitivity values,
as given in Table B.3. of \cite{van13} for a source at 30\degr
declination, to the aforementioned bandwidth and sample time. The
sensitivity value quoted for 150~MHz is in good agreement with the
value from the LPPS survey \citep{coe14} using uncalibrated LOFAR
stations, scaled to this survey.

For this survey, groups of 30 and 31 beamlets are pointed in a single
direction on the sky, covering contiguous frequency channels to form a
total of 8 broadband ($\sim6$~MHz) beams. These can be formed within a
circle of $\sim$10 times the beamlet FWHM, due to the analog tile
beamforming of LOFAR HBA arrays described in \cite{sta11}.

\begin{table}
\caption{\label{tab:specs} Specifications of an international
  LOFAR station}
\centering
\begin{tabular}{@{}ccc}
  \hline 
  Frequency & Beamlet FWHM & Sensitivity \\
  (MHz) & (deg) & (Jy, $\Delta
  \nu$=6~MHz, $\delta t$=327.68~$\upmu$s) \\
  \hline
  \hline
  30 & 9.9 & 1566 \\ 
  60 & 4.0 & 1079 \\ 
  120 & 2.5 & 33\\
  150 & 2 & 27 \\ 
  \hline
\end{tabular}
\begin{tabular}{@{}rl}
  Elevation range: & 30$\degr$ to 90$\degr$\\
  Maximum number of beamlets: & 244 \\ 
  Maximum total sky bandwidth: & 47.65625 MHz\\
  Sampling rate: & 200 or 160 Msamples/s\\
  Beamlet frequency resolution: & 195.3125 or 156.25 kHz \\
  Beamlet time resolution: & 5.12 or 6.4 $\upmu$s\\
  \hline
\end{tabular}
\end{table}

Each ARTEMIS server receives and processes 2 out of the 8 formed
beams, corresponding to a quarter of the total bandwidth.  In addition
to FRB detection, real-time processing can be used to reduce the data
rate from 400~MB/s to manageable rates both for storage and further
processing, useful for other science goals.

\subsection{Software overview and framework}

\begin{figure} 
\includegraphics[width=0.49\textwidth]{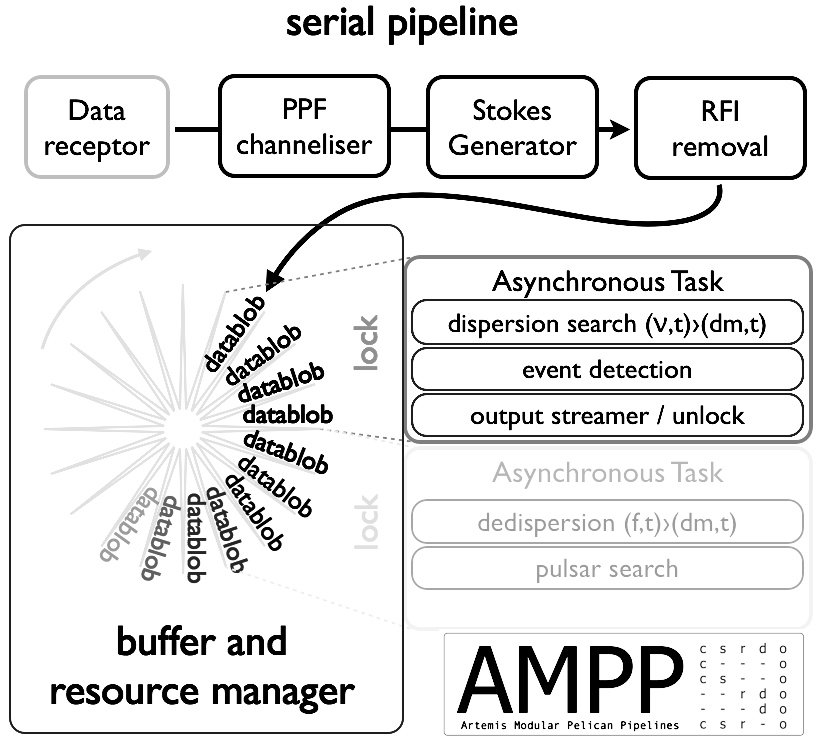}
\caption{A block diagram of the ARTEMIS software, outlining the main
  processing modules, buffers and asynchronous
  tasks. \label{fig:AMPP}}
\end{figure}

ARTEMIS provides an extendible and parallelisable software backend,
which can process telescope data in real time, mainly for detecting
FRBs. The software, called AMPP, is sufficiently generic to serve as
an all-purpose telescope backend, which is built upon a C++ framework
(PELICAN\footnote{\url{http://www.oerc.ox.ac.uk/~ska/pelican/}}),
providing a fully configurable client-server architecture. The
software pipelines consist of a number of C++ modules and
configuration files for processing live telescope data.  The modular
design allows data processing tasks to be implemented independently of
other considerations such as data acquisition, transport protocols,
data output, and system deployment. The modules currently available
include an implementation of a polyphase filter for raw (coherent)
telescope data, a Stokes-generating and integration module which
converts the raw data to Stokes parameters, a radio-frequency
interference (RFI, unwanted and often man-made signals) rejection
module which operates on incoherent data, an incoherent dedispersion
transform module, and streaming modules that record data to disk and
send the processed data via the network for further processing or
recording. Figure~\ref{fig:AMPP} shows a schematic of AMPP. As shown,
dedispersion is an asynchronous task which occurs only when
sufficient data have been accumulated in a buffer. This task is also
the most expensive in terms of computation, and therefore is
accelerated using GPUs. AMPP uses the AstroAccelerate dedispersion
transform algorithm \citep{arm12}, which has been favourably
benchmarked against other extremely fast implementations
\citep{bar12,mag11}.

\subsection{Spectral processing and RFI excision}

In aperture arrays such as LOFAR, beamforming can require that
spectral channels are formed in the raw digital data
\citep{mol11}. This is achieved prior to beamforming via a polyphase
filter followed by a fast Fourier transform, which produces a number
of raw-data frequency subbands. The channelisation requirements of the
beamformer are generally different to those of the dedispersion
transform, therefore an additional channelisation step is necessary
within the software. Incoherent dedispersion cannot remove
intra-channel dispersion. As a result, the channel bandwidth should
not exceed a certain limit, dictated by the maximum DM of a search and
the minimum duration of the FRBs being searched for. Determining this
limit is standard practice in pulsar and fast transient searches
\citep[e.g.,][]{kei10}.

We have developed a multi-threaded polyphase filterbank module for
AMPP, which efficiently generates $2^{\rm n}$ spectral channels for each
input subband. The number of output channels, the windowing function,
the number of taps for the filter and the number of processing threads
dedicated to the module are configurable options. In our searches so
far, we use a filter that provides frequency channels with a bandwidth
of 3.0518 kHz, by channelizing each incoming raw data subband into 64
channels. This increases the sample time from the original 5.12 $\upmu$s
to 327.68 $\upmu$s.

Within the AMPP pipeline, the raw data are passed to a module which
generates Stokes parameters from the incoming complex polarisations,
according to:
\begin{align} 
I &= XX^*+YY^*,& Q &= XX^*-YY^* \nonumber \\
U &= 2 {\rm Re}(XY^*),& V &= -2 {\rm Im}(XY^*),
\end{align}
where $^*$ refers to the complex conjugate.  In the dispersion
searches for FRBs described here, only Stokes-$I$ (total power) is
used. The software however provides the possibility to buffer all four
Stokes parameters.

\begin{figure*} 
\includegraphics[width=0.49\textwidth]{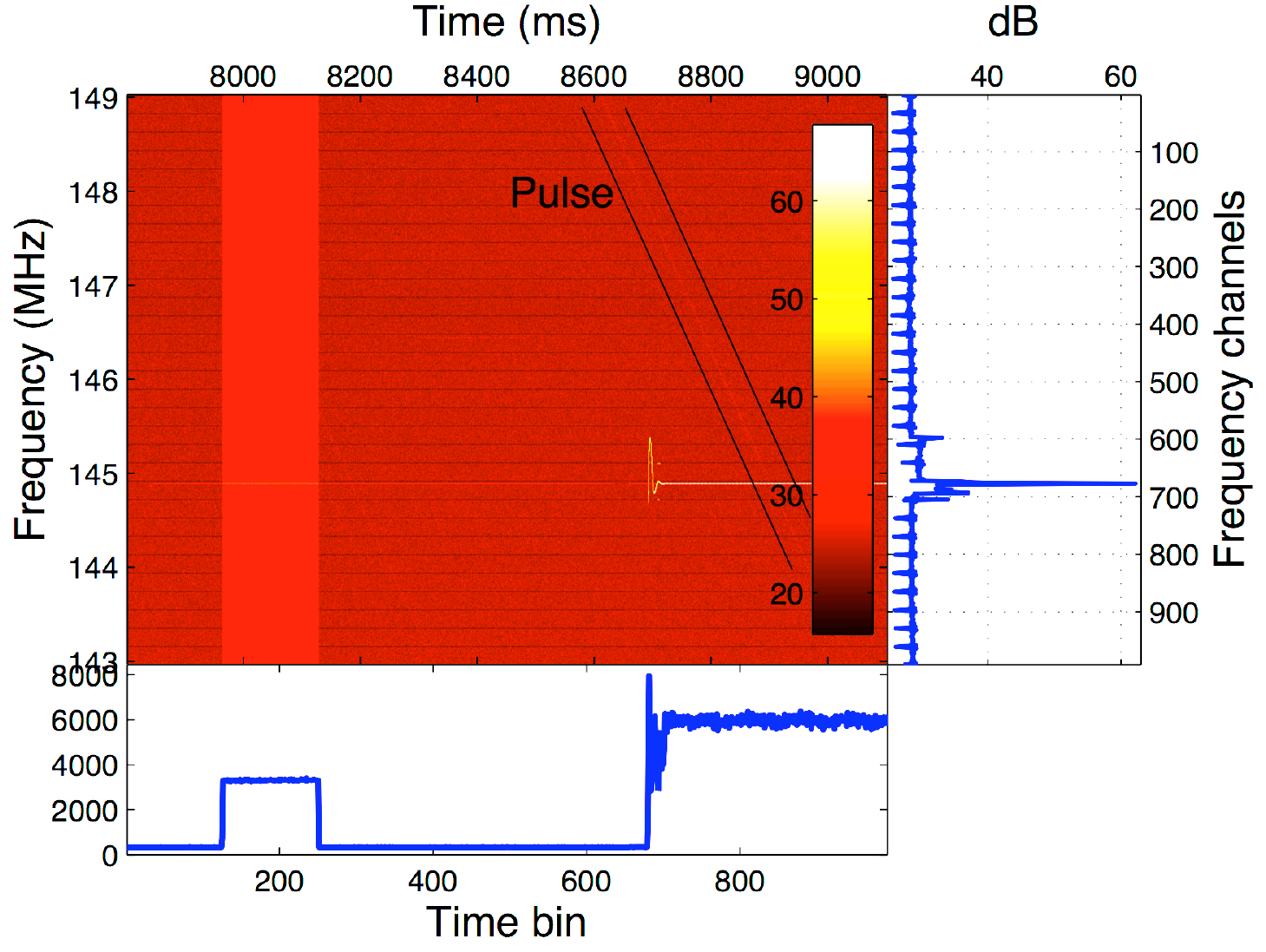}
\includegraphics[width=0.49\textwidth]{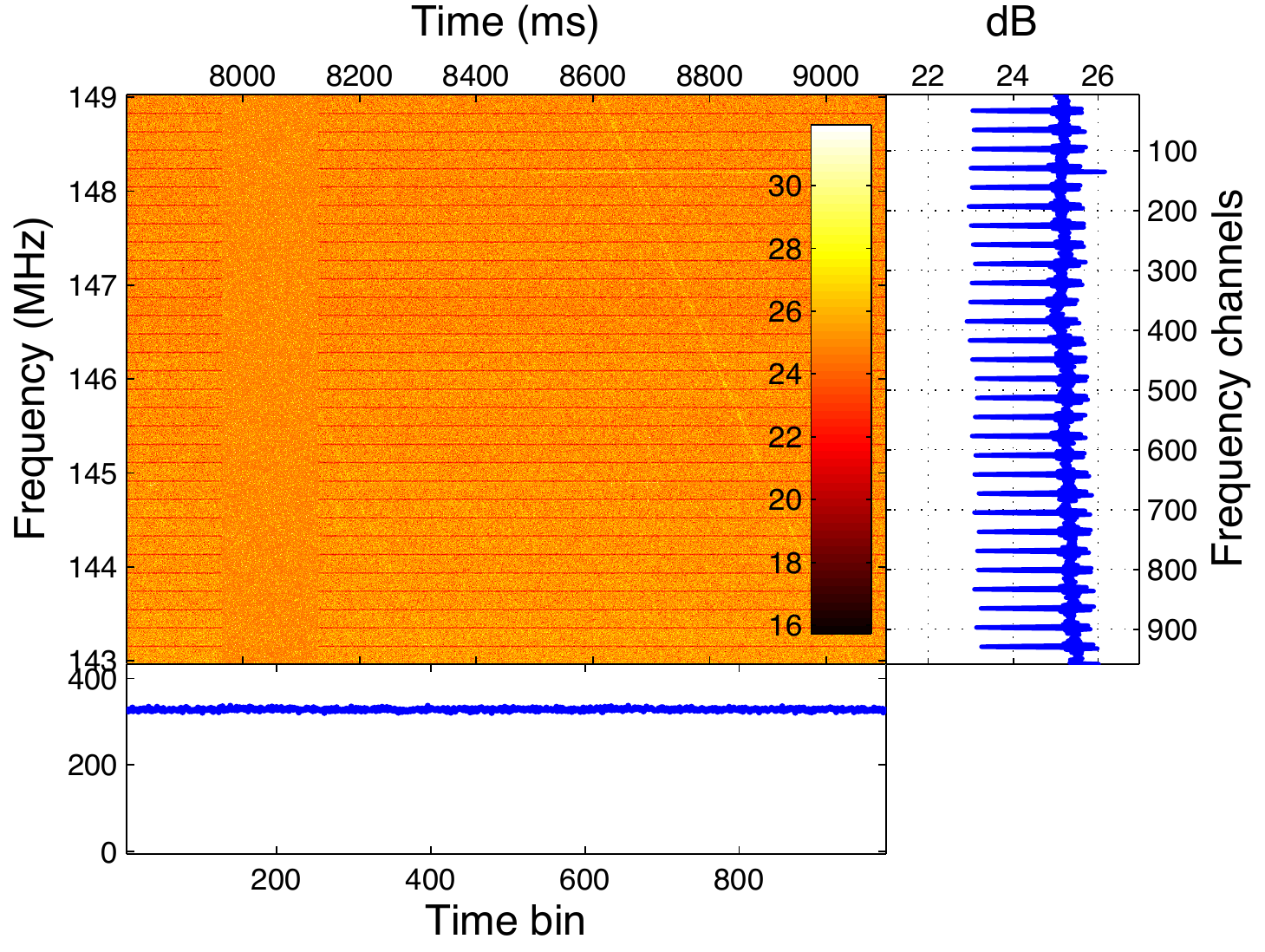}
\caption{Filterbank data from the Rawlings Array. The central images
  show total power versus time and frequency. The bottom panel shows a
  time series of the total power integrated across frequency, and the
  panel to the right shows the bandpass averaged over time. In the
  bandpass, the series of parallel horizontal lines mark the
  boundaries of the LOFAR subbands, which have been further
  channelised in AMPP. The figure on the left shows data before the
  RFI excision module, including narrowband interference at
  $\sim$145~MHz which flares and sweeps across a narrow frequency
  range around time
  bin 680, and a broadband signal around time bin 200. The
  figure on the right shows the same data, post RFI filter. The
  broadband signal at time bin 200 has been flagged and in this case
  replaced by noise. The narrowband interference has been completely
  flagged out, revealing two dispersed bright pulses from PSR B0329+54,
  originally buried in the noise.
  \label{fig:RFI_clipper}}
\end{figure*}

RFI is known to yield large numbers of false positives in pulsar
searches \citep{eat09}. The dedispersion transform integrates total
power over frequency, thereby removing all frequency resolved
information. This imposes the requirement to clean the data of RFI
before dedispersion. Examples of RFI excision techniques for FRB
searches have been described in \cite{hog12}. For our survey, removing
RFI in real-time imposes specific constraints both on the algorithm
(only information from the past data is available) and on the
processing. The adaptive thresholding algorithm that we developed for
AMPP performs the following steps:

\begin{enumerate}
\item Prior to any observation, the spectral bandpass is recorded
  using narrow frequency channels. Individual bright channels are
  rejected through an iterative rejection process, and the clean
  bandpass is modelled using a low-order polynomial. The root mean
  square (RMS) of the bandpass noise is recorded.
\item The search commences.
\item Within AMPP, the total power data are processed by the RFI
  module one time sample at a time, consisting typically of a large
  number of spectral channels. 
\item The median of each time sample spectrum is computed, which is
  insensitive to outliers, and the bandpass model is adjusted to that
  level (the zeroth order coefficient is set to the value of the
  spectrum mean).
\item The power in each frequency channel is compared to the model at
  that frequency. If it exceeds a threshold $T_{\rm ch}$, it is given
  a small weight (in current implementations we set the weight to
  zero). $T_{\rm ch}$ is typically set to 6 times the RMS of the
  current bandpass noise.
\item The mean and RMS of this `cleaned' spectrum are stored in a
  circular history buffer, which can hold values extended up to a duration
  much greater than the longest FRB searched for. 
\item A running average of the history buffer is computed both for the
  model mean and the RMS, ensuring that the model remains up-to-date
  when it is compared to the next observed spectrum, without being
  affected by short term variability.
\item If the median value of the incoming spectrum reveals a large
  disagreement between the model and the spectrum, i.e., a jump has
  occurred in the incoming spectra due to a broadband signal, the whole
  spectrum is assigned a low weight. The threshold $T_{\rm sp}$ is
  typically also set to 6 times the calculated RMS of the median.
\item If the number of consecutive spectra that do not agree with the model
  is equal to the size of the history buffer, the algorithm accepts
  that the bandpass model is no longer relevant to the data, and sets
  the mean and RMS to the values of the last observed spectrum. 
\item Finally, for the FRB search, the bandpass model is removed from
  each flagged spectrum. AMPP has a configuration setting that allows
  each spectrum to be scaled to maintain a uniform RMS in the
  post-filtered data, which is very useful for the dedispersion transform.
\end{enumerate}

Figure \ref{fig:RFI_clipper} shows the effect of the RFI excision
algorithm on 1.3 s of filterbank data from the Rawlings Array. The
algorithm described above effectively flags broadband signals which
raise the mean of the spectrum. In this example, the flagged data are
replaced by Gaussian noise matching the bandpass model used for the
comparison. However, for the real-time FRB searches, we set the weight
of these data to zero, and replace them in the input to the
dedispersion module with noise from a chi-squared distribution with
four degrees of freedom.  A narrowband (and in this case
frequency-swept) signal of interference is also flagged, revealing two
dispersed pulses of PSR B0329+54 in the processed filterbank data on
the right. The RFI excision module can also be set to remove the mean
of every spectrum, effectively removing signals at 0 DM \citep{eat09}.

\subsection{Dedispersion transform considerations}

The dedispersion module used in AMPP was developed as part of the
Astro-Accelerate library \citep{arm12}. This module performs the
dedispersion transform between total power versus frequency and time,
and total power versus dispersion measure and time. The software
framework provides the infrastructure for setting up tasks on
available GPUs, and GPU dedispersion kernels are used during AMPP
processing. The details of the dedispersion process are beyond the
scope of this paper and form the subject of a separate
publication. Together with the RFI-cleaned data, the mean and standard
deviation of the baseline noise are also propagated through the
software pipeline to facilitate the detection of statistically
significant signals in the de-dispersed data.

Incoherent dedispersion is performed in steps of DM. For technical
reasons, pertaining to the dedispersion algorithm on a GPU, we
maintain a fixed $\delta$DM step across the searched DM range.  The
dedispersion transform in our survey was therefore carried out at
$\delta$DM steps of 0.1~cm$^{-3}$ pc. This discretization results in
some loss of signal due to intra-channel dispersion. \cite{cor03}
provide a framework for estimating the signal loss given the DM step
size, the centre frequency and bandwidth of the observations and the
width of a Gaussian pulse. Using their equations 12 and 13
\begin{equation}\label{eq:sovers}
\frac{S(\delta{\rm DM})}{S} = \frac{\sqrt{\pi}}{2} \zeta^{-1} {\rm
    erf}\zeta,
\end{equation} 
where
\begin{equation}\label{eq:zeta}
\zeta = 6.91 \times 10^{-3} \delta{\rm DM} \frac{\Delta\nu_{\rm
    MHz}}{W_{\rm obs}\nu^3_{\rm GHz}}.
\end{equation}
and $\delta{\rm DM} = 0.1$~cm$^{-3}$ pc, we estimate a loss in
sensitivity of less than 15\% for 2~ms wide pulses, which drops to
2.5\% for a 5~ms pulse.


 
Regardless of prior beliefs on the origin of FRBs, at a sky frequency
of 145~MHz, interstellar scattering is a central consideration in the
design of a search. \cite{bha04} compute an empirical relationship
between the measured scattering timescale $\tau$ and the DM of a large
number of pulsars. There is at least an order of magnitude uncertainty
in the DM-$\tau$ relationship. Scattering results in broadening of the
intrinsic FRB width, resulting in a smaller loss of sensitivity, as
per eqs. \ref{eq:sovers} and \ref{eq:zeta}. To investigate relevant
values of W$_{\rm obs}$ we assume the worst case scenario as regards
dedispersion (i.e., the least possible scattering per given DM). We
therefore set $\tau_{\rm min}$ to be an order of magnitude smaller
than what is given by \cite{bha04}. In principle, the observed width
of a radio pulse is given by the convolution of $\tau_{\rm min}$ with
the intrinsic width of the burst, although for simplicity, we derive
the observed width as
\begin{equation}\label{eq:width}
  W_{\rm obs}^2 = W_{\rm int}^2 + \tau_{\rm min}^2,
\end{equation}
where the intrinsic width $W_{\rm int}$ is defined in some reasonable
way (e.g., FWHM). Another factor that comes into play specifically for
signals from cosmological distances is cosmological time dilation,
which stretches the pulse by a factor of $(1 + z)$. However, as
discussed in the following sections, $z$ is small in the case of the
ARTEMIS survey, so this effect can be safely neglected.

\begin{figure} 
\includegraphics[width=0.49\textwidth]{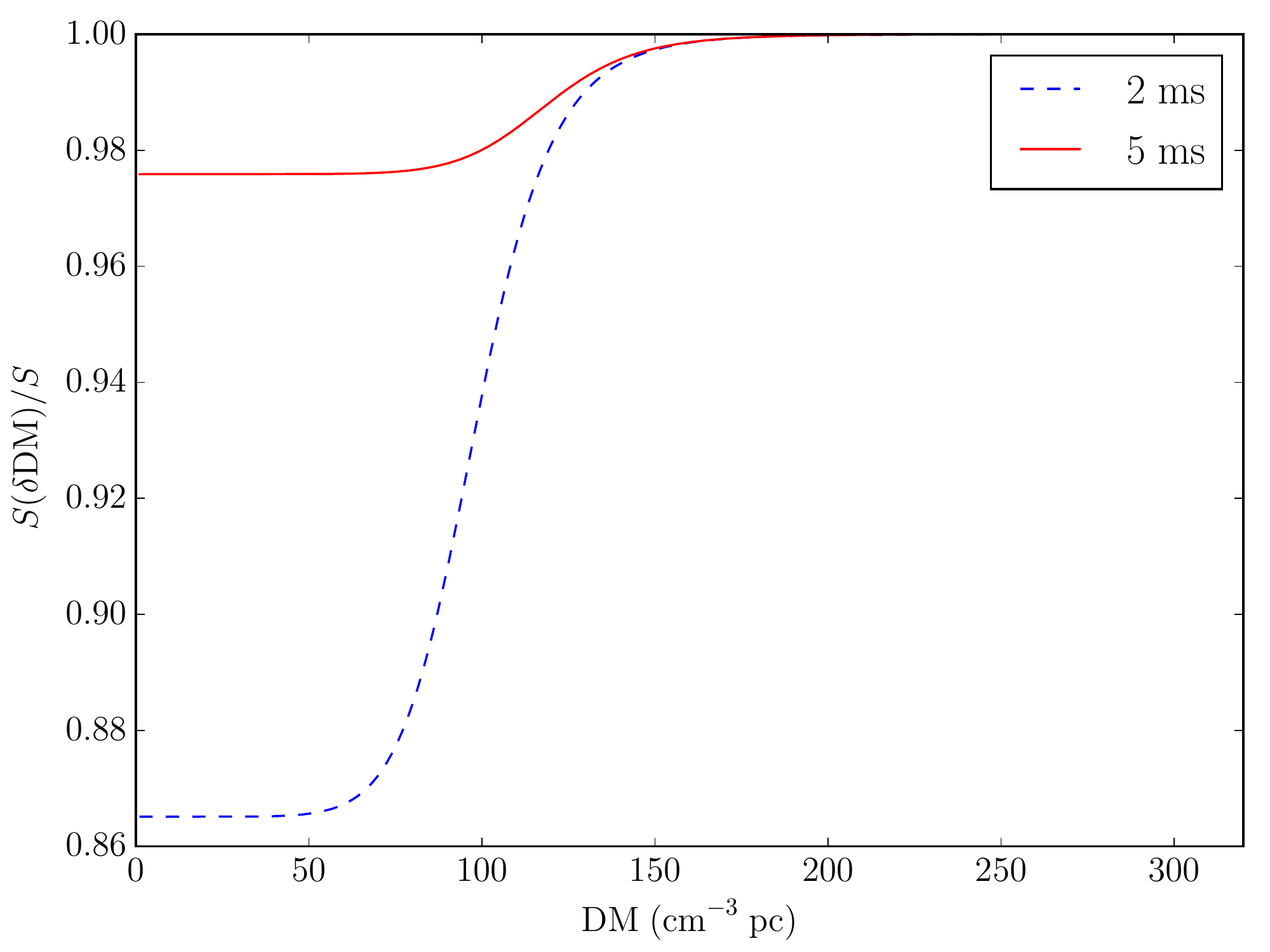}
\caption{The efficiency $\frac{S(\delta{\rm DM})}{S}$ for a
  $\nu=145$~MHz and $\Delta\nu=6$~MHz search for FRBs with intrinsic
  width of 2 and 5~ms, given as a function of DM. The curve is
  computed using Eqs. \ref{eq:sovers}, \ref{eq:zeta}, \ref{eq:width},
  and setting $\tau_{\rm min}$ to be 10\% of what is given by
  \citet{bha04}. Regardless of the intrinsic FRB width, $W_{\rm obs}$
  is likely dominated by the scattering timescale $\tau_{\rm min}$
  above a certain DM, thus rendering the efficiency high even for low
  intrinsic widths.
  \label{fig:DDMvDM}}
\end{figure}

In the LOFAR HBA band at 145~MHz, there is a reasonable expectation
that $W_{\rm obs}$ will be dominated by scattering above a certain DM.
Figure \ref{fig:DDMvDM} shows the efficiency $\frac{S(\delta{\rm
    DM})}{S}$ for a DM resolution $\delta$DM=0.1~cm$^{-3}$ pc, for
FRBs with an intrinsic width of 2 and 5~ms, based on the arguments
above. The curves are characterised by a flat segment at the lowest
DMs, which indicates the DM range for which Galactic scattering is
unlikely to substantially broaden the pulse width ($W_{\rm obs}
\approx W_{\rm int}$). The figure suggests that a DM resolution of
$\delta DM = 0.1$~cm$^{-3}$ pc maintains high efficiency even at low
DM values.

Considering that FRBs typically last a few milliseconds sets an upper
limit to the DM value searched using incoherent dedispersion. The
reason is that intra-channel dispersion becomes large with increasing
DM value at these low frequencies. Reducing the channel bandwidth is
not an option, as it worsens the temporal resolution. For the channel
bandwidth of 3.0518~kHz, the DM smearing at 145~MHz is 8.3$\mu$s per
unit of DM. We have therefore chosen the maximum DM of our search to
be 320~cm$^{-3}$~pc, corresponding to a maximum of 2.6~ms of
intra-channel smearing. To search for higher DM FRBs requires a step
of coherent dedispersion, which would leave the temporal resolution
unaffected.

The ARTEMIS hardware is capable of processing at least 3200 DM values
per data stream in real-time, which is sufficient given the DM
resolution arguments above. In fact, this sampling of the DM space was
also chosen to demonstrate the viability of the hardware and software,
operating at near maximum load. Dedispersed timeseries at neighbouring
DM values will show some degree of correlation, however the threshold
used for detection (10 times the noise RMS, see the following
section), precludes spurious detections caused by this effect.

Once the filterbank data have been converted to a time series per DM,
these are decimated to 2, 4, 8, 16, 32 and 64 times the original
sampling time, and, along with the original, searched in real time for
significant peaks by applying a 5-sample median filter to preclude the
detection of single, high S/N peaks. We are currently investigating
adaptive filtering algorithms for computational and detection
performance.

\section{FRB searches using the Rawlings Array}\label{sec_surv}

We have currently concluded 1522~hours of drift-scan
observations using the Rawlings Array. These observations were
accompaniedfor the majority of the time by the French LOFAR
international station at Nan\c{c}ay. The purpose of the second station
was to perform a coincidence test on potential detected FRBs and rule
out the possibility that they are caused by RFI.

The array is configured to point 8 beams with a bandwidth of 6.055~MHz
(for odd-numbered beams) and 5.859~MHz (for even-numbered beams), to
fixed positions in the sky on the local meridian, at declinations of
7.9\degr, 11\degr, 13.3\degr, 15.85\degr, 19\degr, 22\degr, 24.9\degr
and 28.66\degr, corresponding to declinations of known bright pulsars
which we use as sanity checks. Each beam is about 1.9\degr wide, and
the total amount of instantaneous sky coverage is 24 sq. deg. Our
sensitivity is given by
\begin{equation}
S_{\rm min} = 24.7~{\rm Jy} \left(\frac{10}{\sqrt{D}}\right),
\end{equation}
where 24.7~Jy is the noise RMS in each sample averaged across the
band. This is computed by correcting the value given in
Table~\ref{tab:specs}, by the average elevation of our beams
(58\degr).  Furthermore, 10 is the threshold S/N ratio that we used,
and $D$ is the decimation factor, compared to the native
327.68~$\upmu$s time resolution. This makes us sensitive to a peak
flux density $\gtrsim 62$~Jy for 5~ms-duration pulses. This
corresponds to a fluence, i.e. energy contained within a pulse, of
310~Jy~ms. Note that, given the other decimation factors we have
searched for, our survey has multiple fluence limits, ranging from
81~Jy~ms for D=1 (pulse durations of 327$\upmu$s) and 650~Jy~ms for
D=64 (pulse durations of 21~ms).

\begin{figure} 
\includegraphics[width=0.49\textwidth]{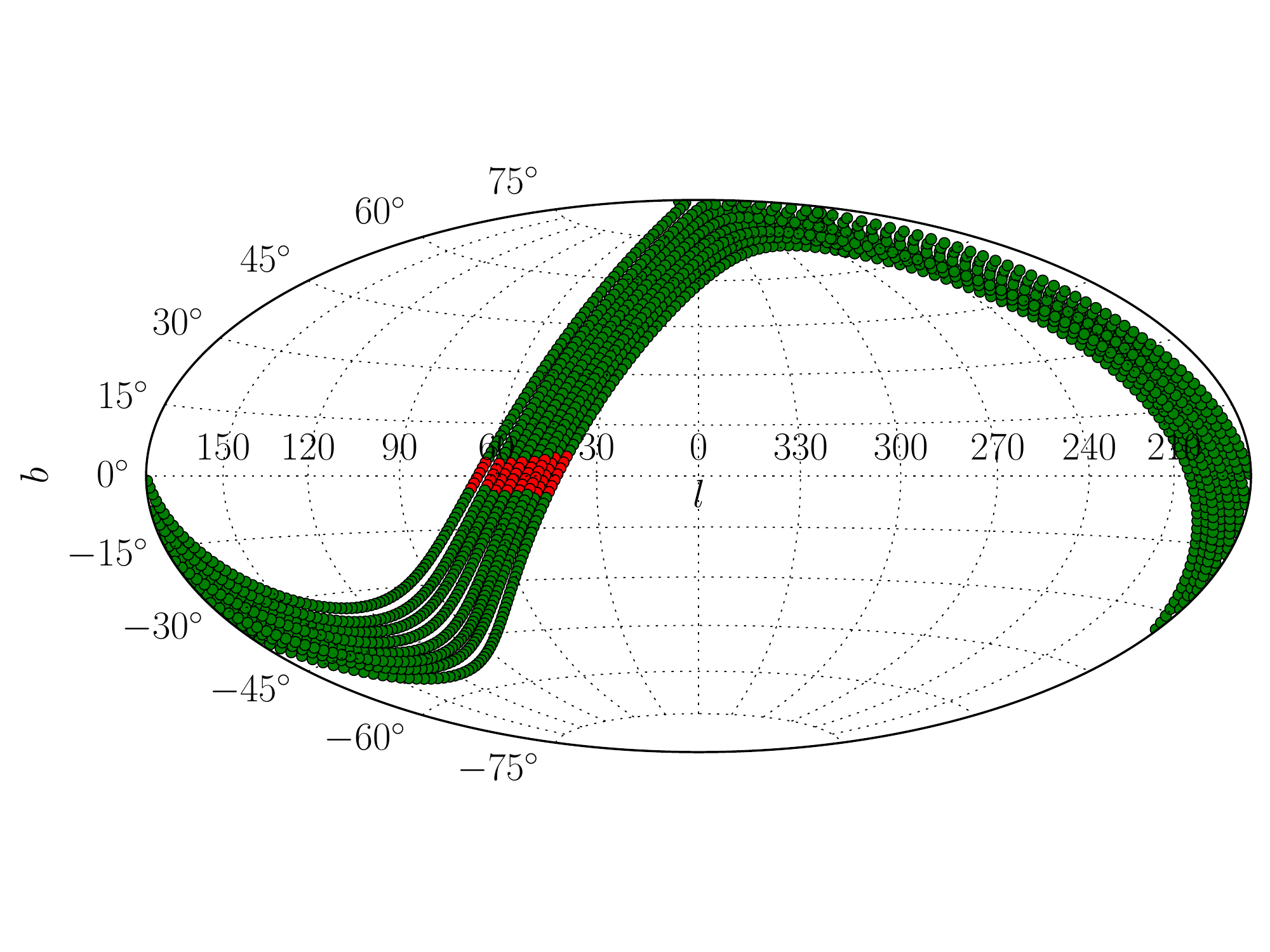}
\caption{The pointings of the eight beams of the drift-scan survey. The
red markers represent locations where the Galactic plane obscures extragalactic
sources, while the green markers represent pointings that probed outside the
Galaxy. The size and shape of the markers are not to scale.\label{fig:vol}}
\end{figure}

The survey searched for FRBs up to a DM of 320~cm$^{-3}$ pc, covering
a volume of $1.15 \times 10^8$~Mpc$^3$, which was computed in the
following manner: Using the NE2001 model of electron density in the
Galaxy \citep{cor02} and making the simple assumption that the
potential host galaxy contributes a constant DM of 100~cm$^{-3}$ pc,
we can estimate the excess DM due to the IGM in each of our
pointings. Figure~\ref{fig:vol} shows the pointings of the survey in
Galactic coordinates. Pointings that did not probe outside the Galaxy
(in the Galactic plane, indicated by red markers in
Figure~\ref{fig:vol}) were excluded from the volume calculation.
Then, following \cite{lor07}, we used the results of \cite{iok03} and
\cite{ino04} to estimate the redshift corresponding to the IGM
component of the DM. There are considerable uncertainties in the
relationship between redshift and DM. Acknowledging this, we proceed
by adopting DM=1200~$z$~cm$^{-3}$ pc.  Assuming a flat universe, we
compute the comoving distance for each of the pointings, and the
volume covered by the beam to that distance. The volumes corresponding
to each extragalactic pointing are then summed to get the total
volume. With the above assumptions, the maximum redshift probed in
this survey was 0.17, and the average redshift among all our pointings
is 0.13.  We stress again the large associated uncertainties.

The average Galactic DM of all pointings that probed extragalactic
space is 61.6~cm$^{-3}$~pc, which, given our considerations about the
host galaxy DM, justifies our DM spacing of 0.1~cm$^{-3}$~pc
(Figure~\ref{fig:DDMvDM}). Given estimates of scatter-broadening
related to these DM values in pulsars \citep{bha04}, and comparatively
small intergalactic scattering \citep{lor13}, we do not a priori
expect scattering to render FRBs undetectable at the frequency and in
the DM range of our survey.

\begin{figure} 
\includegraphics[width=0.49\textwidth]{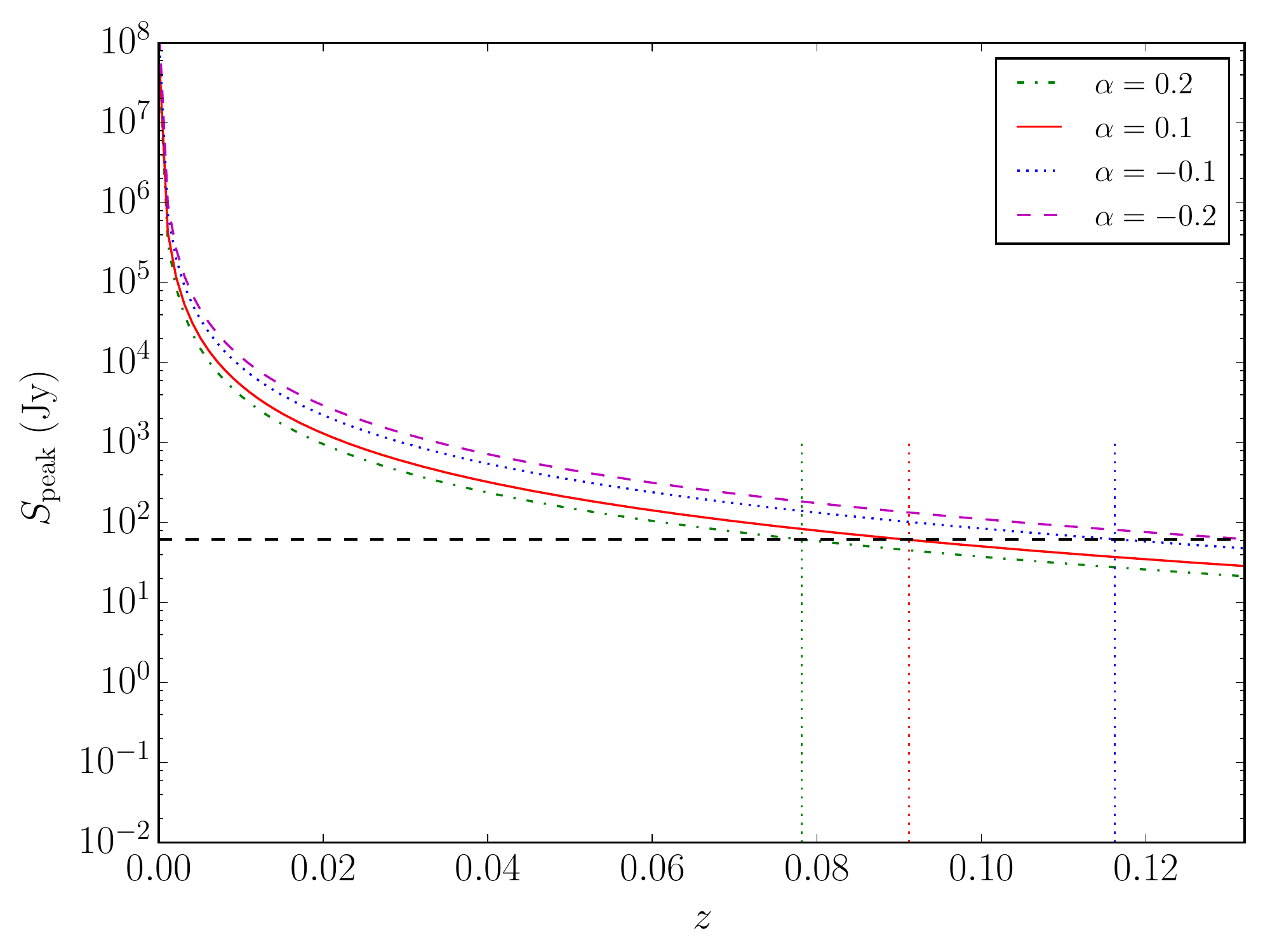}
\caption{The relationship between the observed peak flux density for
5~ms-duration pulses and the source redshift, based on treating FRBs as
standard candles \citep{lor13}, for various values of spectral index $\alpha$.
The black horizontal dashed line is the sensitivity limit of the ARTEMIS
survey for 5~ms-duration bursts. The vertical lines are at the same
redshifts as the vertical dotted lines in Figure \ref{fig:Rz}, and
help provide limits on the estimated spectral index of $\alpha = 0.1$, as described in detail in the text.
\label{fig:Szalpha}}
\end{figure}

Assuming FRBs are standard candles with broadband emission, the volume
sampled by the survey may depend on their spectral index rather than
the maximum DM. This can be understood as follows. Sources at a given
maximum redshift should be bright enough at the frequency of
observation to be detectable.  Following \cite{lor13}, we may be
overestimating the volume in our survey if the spectral index of the
sources is positive, as shown in Figure~\ref{fig:Szalpha}, where eq.~9
from \cite{lor13} is plotted for various values of spectral index.
Figure~\ref{fig:Szalpha} shows the relationship of the peak flux
density of a 5~ms FRB as a function of redshift and spectral
index. Lines of different spectral index -- 0.2, 0.1, $-$0.1,
$-$0.2 -- are plotted. A horizontal line is placed at the sensitivity
limit of the ARTEMIS survey.  For a given maximum redshift, the
minimum spectral index of detectable FRBs is the one that leads to a
curve that intersects the sensitivity line at that redshift.

Figure~\ref{fig:vol} shows the sky coverage of the survey. Out of all
the pointings in the survey, about 5\% of pointings were limited to
within the Galaxy due to the DM limit. We have excluded these
pointings, leading to a total sky coverage in our survey of 4193
sq. deg. The total duration of the survey as it pertains to
sensitivity towards extragalactic bursts reduces to 1446
hours. The single horizontal sensitivity line in
Figure~\ref{fig:Szalpha} is justified by the majority of our pointings
being at high Galactic latitudes ($>$20\degr), where the sky
temperature does not vary significantly.

\begin{figure} 
\includegraphics[width=0.49\textwidth]{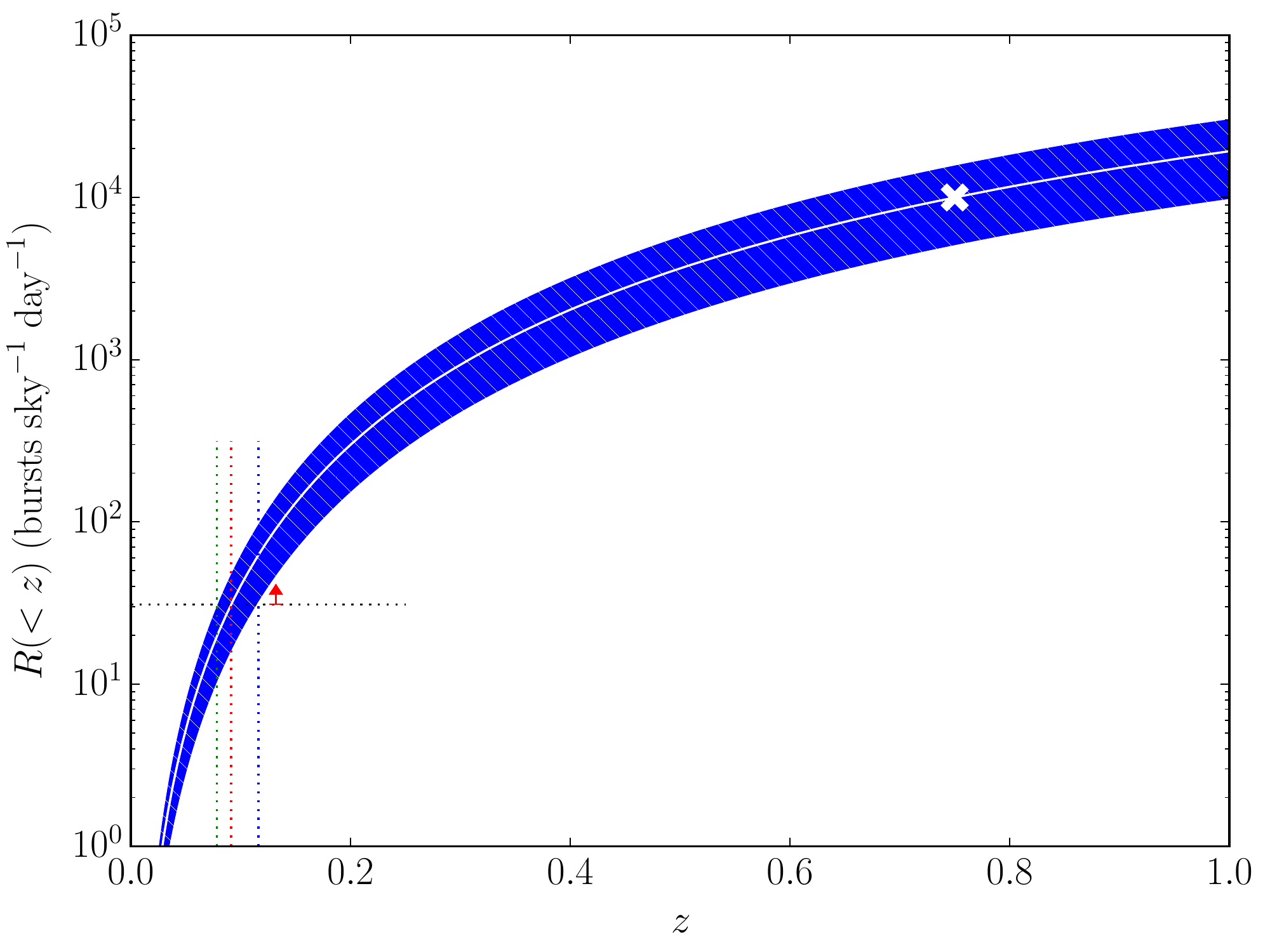}
\caption{Reproduction of the centre panel of Figure~2 from
  \citet{lor13}. The white cross is the \citet{tho13} rate and the red
  arrow is the ARTEMIS upper limit. The white curve represents the FRB
  event rate at all redshifts constrained by the \citet{tho13} rate,
  assuming that their survey was complete up to $z = 0.75$.  The band
  around it represents the Poisson error on the \citet{tho13}
  rate. The horizontal dotted line is at the level of the ARTEMIS
  upper limit, and intersects the extrapolated Thornton rate (and
  errors) at the vertical lines, which correspond to the redshift
  limits shown in Figure \ref{fig:Szalpha}.
  \label{fig:Rz}}
\end{figure}

\subsection{Results}

\begin{figure*} 
\includegraphics[width=\textwidth]{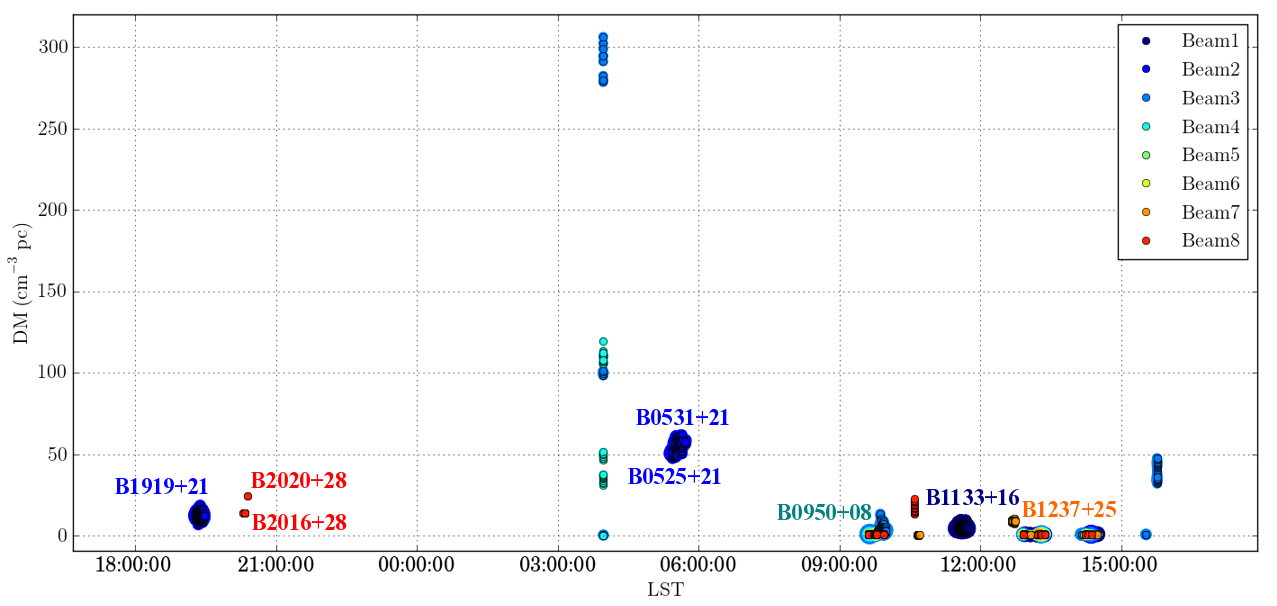}
\caption{A sample output plot of single pulse detections in the
  ARTEMIS survey, showing a few pulsars and some RFI. Each cluster of
  markers correspond to single pulses from a pulsar, while the sets of
  points, elongated in DM, are RFI.
\label{fig:dmvstplot}}
\end{figure*}

Our 1446-hour survey resulted in no detection of
FRBs. Figure~\ref{fig:dmvstplot} shows an example of recorded events
during the observing session of 19 September 2014. These consist of
single pulses from known pulsars and sources of RFI. During the
survey, the real-time processing software detected single pulses from
all expected bright pulsars, including B0950+08, B1919+21, B1133+16,
B0525+21, B0531+21 (the Crab pulsar), B2016+28, B2020+28, and
B1237+25. This demonstrated the functionality of the system, including
the ability to process a large number of DMs in real-time. We also
serendipitously detected signals from HAMSAT, a low-Earth-orbit
satellite whose signals appear at low, non-zero DMs. Despite the
sophisticated adaptive RFI excision algorithm, a small number of
spurious events were recorded per epoch, which were all evaluated by
eye. For a 5 ms resolution, the total number of such events accounts
for approximately 10$^{-4}$ of the total searched parameter space in
DM and time, and is therefore negligible in all calculations.

Based on the sky coverage and the time surveyed, and assuming an
isotropic distribution of sources and completeness of the survey
\citep[for caveats, see][]{bur14,kea15}, we derive an upper limit to
the rate of FRBs at 145~MHz, of 29~sky$^{-1}$ day$^{-1}$ for
5-ms-duration events above 62~Jy.

The non-detection of FRBs at low frequencies and their low numbers at
1.4~GHz makes inferring the nature of these sources difficult. Given
the estimated survey volume, we could have observed FRBs from
1.15$\times$10$^6$ host galaxies, assuming a galaxy density of
10$^{-2}$ Mpc$^{-3}$. The amount of time however that each of these
galaxies was covered by the survey beams is $\sim10^{-3}$~yr, given
the size and declination of the beams.  Assuming the FRB volumetric
rate in \cite{tho13} of 10$^{-3}$~Gal$^{-1}$~yr$^{-1}$, that leaves us
with a single potential FRB event in our survey. We therefore have to
consider the possibility that no FRB occurred during the survey. Given
the uncertainty in the relationship between DM and redshift, and given
the uncertainty in the volumetric rates of FRBs at the present day, it
is useful to consider other possible reasons for the non-detection,
such as the spectral index of FRB emission.

As shown in Figure~\ref{fig:Szalpha}, where eq.~9 of \cite{lor13} is
plotted for various values of the spectral index $\alpha$, and as
mentioned in the previous section, spectral index affects the volume
surveyed.  For values of $\alpha$ less than -0.2, the volume surveyed
is dictated by the DM limit. For these values of $\alpha$, following
the standard candle model, we expect to see all bursts up to $z
\approx 0.13$.

Figure~\ref{fig:Rz} reproduces the centre panel of Figure~2 of
\cite{lor13} up to $z = 1$, where the sky rate of bursts enclosed
within a given volume is shown. The \cite{tho13} rate is marked by the
white cross. Our upper limit on the rate, of 29 bursts
sky$^{-1}$~day$^{-1}$, sits below the curve, and is marked by the red
arrow. Assuming that FRBs occurred during our survey and given the
caveats mentioned above, our null result could be explained by the
most distant FRBs being too weak to observe, due to their
spectrum. This leads to the realization that the maximum redshift is
not set by DM considerations, but rather by the spectral index.  To
estimate the new maximum redshift, we observe that our sky rate is
compatible with the \cite{tho13} rate at a lower maximum redshift of
0.09$^{+0.02}_{-0.01}$. This is the redshift at which the curve of the
scaled Thornton rate of Figure~\ref{fig:Rz} also becomes
29~sky$^{-1}$~day$^{-1}$. Figure~\ref{fig:Szalpha} allows us to
associate a spectral index to the new maximum redshift probed, and
errors on that spectral index based on the errors on the maximum
redshift.  Reading off Figure~\ref{fig:Szalpha}, this line of thinking
allows us to place a lower limit on the spectral index of
0.1$^{+0.1}_{-0.2}$.

We must point out that this line of reasoning also reduces the
effective volume of the survey to 3.3 $\times$ 10$^7$ Mpc$^3$, leaving
approximately one third of the host galaxies we started off with that
could host an FRB. If, however, the true reason for our non-detection
lies in the spectral properties rather than the volumetric rate, the
implication is that the FRB spectral index is distinct from that of
most pulsars. In the case of pulsars, a steep (negative) spectral
index is thought to be indicative of a coherent emission process. The
lack of such a steep spectrum in FRBs would serve as a clue to the
nature of their emission process, which could be intrinsically more
narrow-band than pulsars, or characterized by a frequency dependent
emission geometry. It is worth noting that two of the published FRBs
at this time have been observed to have a positive spectral index
within the band of observation: a most likely value of $\alpha$
between 7 and 11, considering the frequency dependent telescope beam
pattern, in \cite{spi14}, and $1.3\pm1.6$ in \cite{rav14}. Estimating
spectral indices within an observing band is difficult due to effects
of scintillation and frequency dependent gain variation within the
telescope beam. This latter effect is more likely to render the
observed spectrum to have a more negative slope than its intrinsic
slope.  Therefore, despite the small population of FRBs with a
positive spectral index, a consistent picture may be emerging.

In estimating the FRB spectral index from our survey we have assumed
that FRBs are extragalactic standard candles with broadband radio
emission. We have also implicitly assumed that the volumetric rate of
FRBs, that reflects the rate per galaxy per year, is constant with
redshift. We have not tested the standard candle hypothesis, while the
volumetric rate may be a function of redshift, similar to the redshift
dependent star formation rate.

\section{Discussion and conclusions}\label{sec_concl}

We have developed ARTEMIS, a GPU-based real-time fast transient
detection system that works with individual LOFAR stations. We have
used the deployment at the Rawlings Array to demonstrate the
functionality and performance of the system using known pulsars. We
conducted a 1446~hour drift-scan survey for FRBs, searching the
DM space up to 320~cm$^{-3}$ pc. With a series of arguments relating
to sensitivity and the intrinsic spectra of FRBs, and some simple
assumptions about their nature, we estimate that we have surveyed a
volume of $3.3 \times 10^7$~Mpc$^3$.

No FRB was detected in the survey at the Rawlings Array or indeed the
Nan\c{c}ay station, and we derive an upper limit of 29 FRBs~sky$^{-1}$
day$^{-1}$ within the parameters of our search. The non-detection
could be due to one or more of the following reasons: Not enough
volume surveyed for enough time, interstellar and intergalactic
scattering, and the intrinsic spectra of FRBs.

Assuming the latter, we have attempted to constrain the spectral index
of FRBs, to $\alpha \gtrsim 0.1^{+0.1}_{-0.2}$, assuming a standard
candle model. The physical significance of this result is moderated by
the fact that we cannot rule out other causes of non-detection at the
frequency of our survey, including the possibility that FRBs are local
events to the telescopes that have detected them. Assuming that the
non-detection is indeed due to the intrinsic shape of FRB spectra, our
result is constrained by the sensitivity of the Rawlings Array.

This work has implications for future low frequency FRB
surveys. Firstly, at frequencies around 145~MHz, longer duration
surveys with increased telescope sensitivity could allow further
constraints of the spectral index to even higher values, or detect
FRBs. It is however quite difficult to sample substantially more
volume of Universe, without running into greater potential scattering
effects. Secondly, moving up in frequency by a factor of 2 and
increasing the sensitivity by two orders of magnitude, which will
ultimately be interesting for the low frequency component of the
Square Kilometre Array, could allow surveys to probe redshifts upwards
of $z$=1. Assuming scatter broadening is mainly due to propagation in
the host galaxy and our galaxy, there is a reasonable expectation that
it will not affect these predictions. This suggests that future
high-sensitivity surveys at 300~MHz, should result in a
reasonable yield of new FRBs and should resolve the spectral index question.

For ARTEMIS, we are developing a commensal real-time system in
collaboration with the Berkeley SETI Research Center, for the Arecibo
radio telescope.  Named ALFABURST, this system will use the seven-beam
$L$-band feed of the Arecibo telescope to perform surveys for FRBs,
potentially increasing the sample size. In this system, we aim to
incorporate multi-beam coincidence rejection that will reduce the
number of false alarms, and automated triggering that will alert
low-frequency telescopes in the event that an FRB is detected. Similar
systems are being prepared for the South African SKA pathfinders
(KAT-7 and MeerKAT) and for tied-array beams of the LOFAR
International Telescope that offer an order of magnitude improvement
in sensitivity.

\section*{Acknowledgments}

We would like to thank Dr Alex Kraus and staff at the Effelsberg
Radiotelescope for hosting us and our machines during early ARTEMIS
development. AK, WA and JC would like to thank the Leverhulme Trust
for supporting this work. The Rawlings Array is operated by LOFAR-UK
as part of the International LOFAR Telescope, and is funded by
LOFAR-UK and STFC. SO is supported by the Alexander von Humboldt
Foundation.


\begin{thebibliography}{}

\bibitem[\protect\citeauthoryear{{Armour} et~al.}{{Armour} et~al.}{2012}]{arm12}
{Armour} W.,  et~al.,
  2012, in {Ballester} P.,  {Egret} D.,   {Lorente} N.~P.~F.,  eds,
  Astronomical Data Analysis Software and Systems XXI Vol.~461 of Astronomical
  Society of the Pacific Conference Series, {A GPU-based Survey for Millisecond
  Radio Transients Using ARTEMIS}.
p.~33

\bibitem[\protect\citeauthoryear{{Bannister} \& {Madsen}}{{Bannister} \&
  {Madsen}}{2014}]{ban14}
{Bannister} K.~W.,  {Madsen} G.~J.,  2014, MNRAS, 440, 353

\bibitem[\protect\citeauthoryear{{Barsdell}, {Bailes}, {Barnes} \&
  {Fluke}}{{Barsdell} et~al.}{2012}]{bar12}
{Barsdell} B.~R.,  {Bailes} M.,  {Barnes} D.~G.,    {Fluke} C.~J.,  2012,
  MNRAS, 422, 379

\bibitem[\protect\citeauthoryear{{Bhat}, {Cordes}, {Camilo}, {Nice} \&
  {Lorimer}}{{Bhat} et~al.}{2004}]{bha04}
{Bhat} N.~D.~R.,  {Cordes} J.~M.,  {Camilo} F.,  {Nice} D.~J.,    {Lorimer}
  D.~R.,  2004, ApJ, 605, 759

\bibitem[\protect\citeauthoryear{{Burke-Spolaor} \&
  {Bannister}}{{Burke-Spolaor} \& {Bannister}}{2014}]{bur14}
{Burke-Spolaor} S.,  {Bannister} K.~W.,  2014, ApJ, 792, 19

\bibitem[\protect\citeauthoryear{Coenen et 
al.}{2014}]{coe14} Coenen T., et al., 2014, A\&A, 570, A60 

\bibitem[\protect\citeauthoryear{{Cordes} \& {Lazio}}{{Cordes} \&
  {Lazio}}{2002}]{cor02}
{Cordes} J.~M.,  {Lazio} T.~J.~W.,  2002, astro-ph/0207156

\bibitem[\protect\citeauthoryear{{Cordes} \& {McLaughlin}}{{Cordes} \&
  {McLaughlin}}{2003}]{cor03}
{Cordes} J.~M.,  {McLaughlin} M.~A.,  2003, ApJ, 596, 1142

\bibitem[\protect\citeauthoryear{{Eatough}, {Keane} \& {Lyne}}{{Eatough}
  et~al.}{2009}]{eat09}
{Eatough} R.~P.,  {Keane} E.~F.,    {Lyne} A.~G.,  2009, MNRAS, 395, 410

\bibitem[\protect\citeauthoryear{{Falcke \& Rezzolla}}{{Falcke \& Rezzolla}}{2014}]
{fal14} Falcke H., Rezzolla L., 2014, A\&A, 562, A137

\bibitem[\protect\citeauthoryear{{Hogden}, {Vander Wiel}, {Bower}, {Michalak},
  {Siemion} \& {Werthimer}}{{Hogden} et~al.}{2012}]{hog12}
{Hogden} J.,  {Vander Wiel} S.,  {Bower} G.~C.,  {Michalak} S.,  {Siemion} A.,
    {Werthimer} D.,  2012, ApJ, 747, 141

\bibitem[\protect\citeauthoryear{{Inoue}}{{Inoue}}{2004}]{ino04}
{Inoue} S.,  2004, MNRAS, 348, 999

\bibitem[\protect\citeauthoryear{{Ioka}}{{Ioka}}{2003}]{iok03}
{Ioka} K.,  2003, ApJ, 598, L79

\bibitem[\protect\citeauthoryear{Keane 
\& Petroff}{2015}]{kea15} Keane E.~F., Petroff E., 2015, MNRAS, 447, 2852 

\bibitem[\protect\citeauthoryear{{Keane}, {Stappers}, {Kramer} \&
  {Lyne}}{{Keane} et~al.}{2012}]{kea12}
{Keane} E.~F.,  {Stappers} B.~W.,  {Kramer} M.,    {Lyne} A.~G.,  2012, MNRAS,
  425, L71

\bibitem[\protect\citeauthoryear{{Keith} et~al.}{{Keith}
  et~al.}{2010}]{kei10}
{Keith} M.~J.,  et~al.,  2010, MNRAS, 409, 619

\bibitem[\protect\citeauthoryear{{Loeb}, {Shvartzvald} \& {Maoz}}{{Loeb}
  et~al.}{2014}]{loe14}
{Loeb} A.,  {Shvartzvald} Y.,    {Maoz} D.,  2014, MNRAS, 439, L46

\bibitem[\protect\citeauthoryear{{Lorimer}, {Bailes}, {McLaughlin}, {Narkevic}
  \& {Crawford}}{{Lorimer} et~al.}{2007}]{lor07}
{Lorimer} D.~R.,  {Bailes} M.,  {McLaughlin} M.~A.,  {Narkevic} D.~J.,
  {Crawford} F.,  2007, Science, 318, 777

\bibitem[\protect\citeauthoryear{{Lorimer}, {Karastergiou}, {McLaughlin} \&
    {Johnston}}{{Lorimer} et~al.}{2013}]{lor13}
{Lorimer} D.~R.,  {Karastergiou} A.,  {McLaughlin} M.~A.,  {Johnston} S.,
2013, MNRAS, 436, L5

\bibitem[\protect\citeauthoryear{Lorimer \& Kramer}{2005}]{lor05}
  Lorimer D.~R., Kramer M., 2004, Handbook of pulsar
  astronomy. Cambridge observing handbooks for research astronomers,
  Vol. 4. Cambridge, UK: Cambridge University Press, 2004

\bibitem[\protect\citeauthoryear{Macquart et 
al.}{2015}]{mac15} Macquart J.-P., et al., 2015, arXiv, 
arXiv:1501.07535

\bibitem[\protect\citeauthoryear{{Magro} et al.}{{Magro} et al.}{2011}]{mag11}
{Magro} A., {Karastergiou} A., {Salvini} S., {Mort} B., {Dulwich} F.,
{Zarb Adami} K., 2011, MNRAS, 417, 2642

\bibitem[\protect\citeauthoryear{{McLaughlin} \& {Cordes}}{{McLaughlin} \&
  {Cordes}}{2003}]{mcl03}
{McLaughlin} M.~A.,  {Cordes} J.~M.,  2003, ApJ, 596, 982

\bibitem[\protect\citeauthoryear{{McLaughlin} et~al.}{{McLaughlin} et~al.}{2006}]{mcl06}
{McLaughlin} M.~A.,  et~al.,
   2006, Nature, 439, 817

\bibitem[\protect\citeauthoryear{{McQuinn}}{{McQuinn}}{2014}]{mcq14} 
McQuinn M., 2014, ApJ, 780, L33

\bibitem[\protect\citeauthoryear{{Mol} \& {Romein}}{{Mol} \&
  {Romein}}{2011}]{mol11}
{Mol} J.~D.,  {Romein} J.~W.,  2011, ArXiv e-prints

\bibitem[\protect\citeauthoryear{{Mottez} \& {Zarka}}{{Mottez} \&
  {Zarka}}{2014}]{mot14}
{Mottez} F.,  {Zarka} P.,  2014, A\&A, 569, A86

\bibitem[\protect\citeauthoryear{{Persic} \& {Salucci}}{{Persic} \&
  {Salucci}}{1992}]{per92}
{Persic} M.,  {Salucci} P.,  1992, MNRAS, 258, 14P

\bibitem[\protect\citeauthoryear{Petroff et 
al.}{2015}]{pet15} Petroff E., et al., 2015, MNRAS, 447, 246

\bibitem[\protect\citeauthoryear{{Popov} \& {Postnov}}{{Popov} \&
  {Postnov}}{2013}]{pop13}
{Popov} S.~B.,  {Postnov} K.~A.,  2013, ArXiv e-prints

\bibitem[\protect\citeauthoryear{Ravi, Shannon, 
\& Jameson}{2015}]{rav14} Ravi V., Shannon R.~M., Jameson A., 2015, ApJ, 799, LL5 

\bibitem[\protect\citeauthoryear{{Spitler} et~al.}{{Spitler} et~al.}{2014}]{spi14}
{Spitler} L.~G.,  et~al.,  2014, ApJ, 790, 101

\bibitem[\protect\citeauthoryear{{Stappers} et~al.}{{Stappers} et~al.}{2011}]{sta11}
{Stappers} B.~W.,  et~al.,  2011, A\&A, 530, A80

\bibitem[\protect\citeauthoryear{{Thornton} et~al.}{{Thornton} et~al.}{2013}]{tho13}
{Thornton} D.,  et~al.,  2013, Science, 341, 53

\bibitem[\protect\citeauthoryear{{Totani}}{{Totani}}{2013}]{tot13}
{Totani} T.,  2013, PASJ, 65, L12

\bibitem[\protect\citeauthoryear{Trott, Tingay, \&
    Wayth}{2013}]{tro13} Trott C.~M., Tingay S.~J.,
  Wayth R.~B., 2013, ApJ, 776, L16

\bibitem[\protect\citeauthoryear{{van Haarlem} et~al.}{{van Haarlem}
    et~al.}{2013}]{van13}
{van Haarlem} M.~P.,  et~al.,  2013, A\&A, 556, A2

\bibitem[\protect\citeauthoryear{{Wayth}, {Tingay}, {Deller}, {Brisken},
  {Thompson}, {Wagstaff} \& {Majid}}{{Wayth} et~al.}{2012}]{way12}
{Wayth} R.~B.,  {Tingay} S.~J.,  {Deller} A.~T.,  {Brisken} W.~F.,  {Thompson}
  D.~R.,  {Wagstaff} K.~L.,    {Majid} W.~A.,  2012, ApJ, 753, L36

\end{thebibliography}

\label{lastpage}

\end{document}